\documentclass[doublecol]{epl2}
\usepackage{bm}
\usepackage{amsfonts}
\usepackage{float}
\usepackage{lscape}
\usepackage{graphicx,color}
\newcommand{\beq}{\begin{equation}}
\newcommand{\eeq}{\end{equation}}
\newcommand{\beqa}{\begin{eqnarray}}
\newcommand{\eeqa}{\end{eqnarray}}
\newcommand{\nn}{\nonumber \\}

\newcommand{\ch}{\mathrm{ch}}

\newcommand{\D}{\Delta}

\newcommand{\disk}{\mathrm{disk}}
\newcommand{\e}{\mathrm{e}}

\newcommand{\la} {\langle}
\newcommand{\ra} {\rangle}

\newcommand{\s}{\sigma}
\newcommand{\z}{\zeta}

\renewcommand{\Im}{\mathrm{Im} \, }
\newcommand{\mod} {\ \mathrm{mod} \ }
\renewcommand{\H}{{\mathcal H}}

\newcommand{\Z}{{\mathbb Z}}

\title{Thermal broadening of the Coulomb blockade peaks in quantum Hall interferometers}

\author{Lachezar S. Georgiev}

\institute{Institute for Nuclear Research and Nuclear Energy, Bulgarian Academy of Sciences, 
72 Tsarigradsko Chaussee, 1784 Sofia, Bulgaria, EU}

\abstract{We demonstrate that the differential magnetic susceptibility of a fractional quantum Hall disk, representing a 
Coulomb island in a Fabry--Perot interferometer, is exactly proportional to the island's conductance and its 
paramagnetic peaks are the equilibrium counterparts of the Coulomb blockade conductance peaks. Using as 
a thermodynamic potential the partition functions of the edge states' effective conformal field theory we find 
the positions of the Coulomb blockade peaks, when the area of the island is varied, the modulations of the 
distance between them as well as the thermal decay and broadening of the peaks when  temperature is increased.
{The finite-temperature estimates of the peak's heights and widths could give important information about the 
experimental observability of the Coulomb blockade. In addition, the predicted peak asymmetry and displacement at 
finite temperature due to neutral multiplicities could serve to distinguish different fractional quantum Hall states 
with similar zero-temperature Coulomb blockade patterns. }
}

\begin{document} 


\maketitle



\begin{figure}[htb]
\centering
\includegraphics[bb=20 200 550 630,width=7cm,clip]{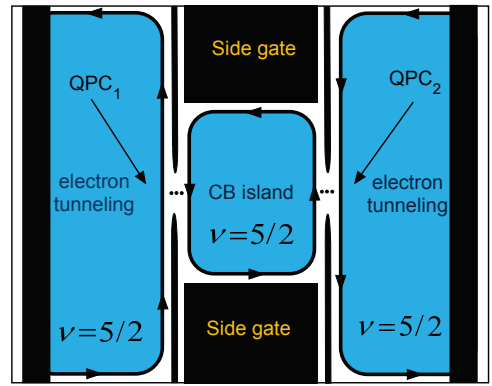}
\caption{(color online) Strong backscattering regime in Fabry--Perot interferometer. Two pinched-off QPCs inside of a 
$\nu=5/2$ FQH bar create a CB island whose area, respectively magnetic flux, could be varied by changing the 
voltage of the side gates.
Electrons could tunnel through the CB island only for special values of the flux for which single-electron spectrum of the island 
is degenerate.}
\label{fig:FPI-CB}
\end{figure}

The structure of the Coulomb blockade (CB) peaks in the conductance of a Fabry--Perot interferometer
\cite{CB-vs-AB,nayak-NA-interferometer}, 
realized by two quantum point contacts (QPC) inside of a fractional quantum Hall (FQH) bar, 
operating in the strong-backscattering regime, depicted in Fig.~\ref{fig:FPI-CB}, which is the stable fixed point 
of the renormalization group (RG) flow, has been widely investigated 
\cite{stern-halperin-5-2,stern-CB-RR,stern-CB-RR-PRB} because of its potential to unveil important intrinsic 
characteristics of the corresponding quasiparticle excitations. However, the zero-temperature CB patterns
appeared to be unable to decisively distinguish FQH states with different topological order
\cite{nayak-doppel-CB,CB,cappelli-viola-zemba}.
In this Letter we will demonstrate how the effective conformal field theory (CFT) for the FQH edges could be 
employed for the computation of the CB peaks' parameters at non-zero temperature
 below the energy gap.
We will show that  the periodicity of the CB peaks could change with increasing the temperature $T$, 
and this could in principle be used to distinguish between different states with identical zero-temperature CB patterns
\cite{nayak-doppel-CB}.

We will be interested in the low-temperature, low-bias transport through the island, formed by the two  
pinched-off QPCs, see Fig.~\ref{fig:FPI-CB}, 
when the leading conductance 
contribution comes from  single-electron tunneling, {known also as sequential tunneling,
 which is the dominant tunneling mechanism close to the center of the CB peaks at low temperature and bias \cite{furusaki}.
 Away from the CB peaks, in the so-called CB valleys, the dominating tunneling process  is called co-tunneling, which is 
 an elastic or inelastic tunneling via virtual intermediate states \cite{averin-nazarov,epl-CB,epl-CB-2,furusaki}, that is responsible for the 
Kondo effect in the Fermi liquid, however in this Letter we will focus on sequential 
tunneling, since we are mostly interested in the modulation of the CB peaks.
The sequential tunneling results from the sequence of  three events:} $A=\left\{\right.$one electron  tunnels from 
the left FQH liquid through the left QPC  to the island$\left. \right\}$,  $B=\left\{\right.$an electron could be accommodated 
at {(and therefore transported along)}  the edge of the CB island$\left.\right\}$ and   
$C=\left\{\right.$an electron tunnels through the right QCP  to the 
right FQH liquid$\left. \right\}$. 
According to the Landauer formula the conductance $G_{\mathrm{CB}} =(e^2/h) P(A \cap B \cap C)$ of the 
Fabry--Perot interferometer in the CB blockade regime is proportional  to the three-event joint probability 
which is a product $P(A \cap B \cap C) =P(A\cap C) P(B)$ because $B$ is statistically independent of $A$ and $C$.
Furthermore, the joint probability $P(A\cap C) =(h/e^2) G_{LR}$ is simply proportional to the conductance of two 
resistors in series $(G_{LR})^{-1}=G_L^{-1}+G_R^{-1}$, where $G_L$ and $G_R$ are the conductances of the 
left and right quantum point contact. Therefore the CB conductance is
\beq \label{G_CB}
G_{\mathrm{CB}} = \frac{g_L g_R}{g_L+g_R} G_{\mathrm{is}} , \quad g_{L,R}= \frac{h}{e^2}G_{L,R} , 
\eeq 
 with $G_{\mathrm{is}}$ being the conductance of the island's edge.
The conductances $G_{L,R}$ are independent of the area of the island and at low 
voltage depend on the temperature as $G_{L,R}\propto T^{4\D -2}$, where $\D$ is the 
scaling dimension of the electron. 

The conductance $G_{\mathrm{is}}$  could be computed at zero temperature
as the derivative of the persistent current, i.e., as a second derivative of the ground state energy, with respect to 
the Aharonov--Bohm (AB) flux $\phi$  using Kohn's relation \cite{Kohn-1964,Imry}
 \beq
\lim_{\omega \to 0} \omega \, \Im \s(\omega)= - \frac{e^2 L^2}{\mathrm{Vol} \, \hbar^2} \, \frac{\partial^2 E_0}{\partial \phi^2}.
\eeq
This remarkable relation can be generalized for non-zero temperatures: 
the conductivity of the edge channel can be written in terms of the charge stiffness (related to the isothermal compressibility) 
due to the Einstein relation (see e.g., Eq.~(2.81) in \cite{DiVentra}),
\beq \label{einstein}
\s(0)=e^2 D \left. \frac{\partial n}{\partial \mu} \right|_{T},
\eeq
where $\mu$ is the chemical potential, $D$ is the diffusion coefficient and the thermodynamic derivative is
taken at constant temperature. The Einstein relation (\ref{einstein}) is proven by the standard Kubo linear 
response  techniques \cite{DiVentra}.

In this Letter we will use the CFT  \cite{CFT-book}  for the FQH edge \cite{wen,fro-zee} to 
define the partition function for the edge and compute the thermodynamic derivative in (\ref{einstein}).

The standard grand-canonical partition function for a FQH disk \cite{kubo}, \cite{CFT-book}, 
\cite{cz,cappelli-viola-zemba,CB}
could be written  in terms of the Bolzmann factor $q=\e^{2\pi i \tau}= \e^{-\D \epsilon/k_B T}$, where the non-interacting 
energy spacing the edge is $\D \epsilon=\hbar \, 2\pi v_F/L$, as  
\beq \label{Z_disk}
Z_{\disk}(\tau,\z) = \mathrm{tr}_{ \H_{\mathrm{edge}}} \ \e^{2\pi i \tau (L_0 -c/24)} \e^{2\pi i \z J_0},
\eeq
where $v_F$ is the Fermi velocity of the edge, $L$ is the edge circumference,  
$H=\hbar\frac{2\pi v_F}{L} \left(L_0-\frac{c}{24}\right)$ is the edge 
Hamiltonian expressed in terms of the zero mode $L_0$ of the Virasoro stress-energy tensor, $c$ is the central 
charge of the Virasoro algebra, $J_0=N$ is the zero mode of the electric $u(1)$ current \cite{NPB-PF_k}, which 
is equal to the Luttinger-liquid particle number operator.
 The Hilbert space $\H_{\mathrm{edge}}$, over which the trace in (\ref{Z_disk}) is taken, corresponds to a single FQH 
edge, i.e.,  the edge of the island which might contain non-trivial quasiparticles in the bulk. The two \textit{modular} parameters \cite{CFT-book}, $\tau$ and $\z$, are purely  imaginary and are related 
to the temperature $T$ and chemical potential $\mu$  as follows \cite{NPB-PF_k}
\beq\label{modular}
\tau=i\pi\frac{T_0}{T}, \ \ T_0=\frac{\hbar v_F }{\pi k_B L}, \quad \z=\frac{ -i }{2\pi  k_B T} \mu .
\eeq
It is worth mentioning that the CFT partition functions are explicitly known for all FQH universality classes.
The FQH disk is threaded by homogeneous perpendicular magnetic field, however, because the dynamics  of the 
FQH liquid is concentrated at the edge,  all thermodynamic quantities depend only on the product of the 
magnetic field and the area of the FQH disk, which could be varied by changing the voltage of the side gate 
\cite{stern-CB-RR}.
The CFT partition function (\ref{Z_disk}) in presence AB flux is modified by  the flux-threading 
transformation \cite{NPB-PF_k}
\beq \label{phi}
\z \to \z +\phi \tau, \quad Z_{\disk}^{\phi}(\tau,\z ) = Z_{\disk}(\tau,\z +\phi\tau), \quad 
\eeq
and is different from the area-variation proposal of Ref.~\cite{cappelli-viola-zemba}.
The physical interpretation of (\ref{phi}) will be discussed after Eq.~(\ref{shift}) below.
Because of this relation between the modular parameters $\z$, $\tau$ and the AB flux $\phi$, we will see below that 
the charge stiffness could be expressed as the second derivative 
of the grand potential of the edge with respect to the AB flux.

In order to compute the particle number average and its derivative with respect to the chemical potential we use 
Eq.~(\ref{modular}) and the standard thermodynamic identification \cite{kubo} of the grand potential  on the edge
$\Omega(T,\mu)=-k_B T \ln Z_{\disk}(\tau,\z)$, 
\beq \label{density}
\la n \ra_{\beta,\mu} = -\frac{k_B T}{L} \frac{\partial}{\partial \mu} \ln Z_{\disk}(\tau,\z)=
\frac{1}{L} \la J_0\ra_{\beta,\mu},
\eeq
where $\beta=(k_B T)^{-1}$ and the thermal average of $A$ is
\beq
\la A \ra_{\beta,\mu} = Z_{\disk}^{-1}(\tau,\z) \, \mathrm{tr}_{ \H_{\mathrm{edge}}} \  A\, \e^{2\pi i \tau (L_0 -c/24)} \e^{2\pi i \z J_0} .
\eeq

Next, in order to apply the Einstein relation (\ref{einstein}) for  $G_{\mathrm{is}}$, we
differentiate the particle density (\ref{density}) with respect to $\mu$ obtaining
\beq \label{stiffness}
\left\la \frac{\partial n}{\partial \mu} \right\ra_{\beta,\mu} = \frac{1 }{L k_B T} \left(\la J_0^2\ra_{\beta,\mu} -(\la J_0\ra_{\beta,\mu})^2 \right).
\eeq
The grand potential $\Omega(T,\mu)$ depends on the AB flux $\phi$ because of (\ref{modular}) and (\ref{phi}). Computing the second derivative  
$\partial^2 \Omega /\partial\phi^2=-(hv_F/L)^2  \left(\la J_0^2\ra -(\la J_0\ra)^2 \right)/k_B T$ and comparing with (\ref{stiffness})
we obtain the main result in this Letter that the conductance $G_{\mathrm{is}}=\s_{\mathrm{is}}(0)/L$ of the edge is 
simply proportional (within Kubo's linear response theory) to the differential magnetic 
susceptibility $\kappa(T,\phi)=(e/h)\partial I(T,\phi) /\partial \phi$, where $I(T,\phi)=-(e/h)\partial \Omega(T,\phi) /\partial \phi$ 
is the persistent current (or, the orbital magnetization) on the edge, i.e.,
\beq \label{susc}
G_{\mathrm{is}}=  \frac{ D }{v_F^2} \kappa(T,\phi), \quad
\kappa=- \left(\frac{e}{h} \right)^2 \frac{\partial^2 \Omega(T,\phi)}{\partial \phi^2}.
\eeq
As we will see below, the CB peaks of the DC conductance $G_{\mathrm{is}}$ correspond precisely to the
paramagnetic peaks of the differential magnetic susceptibility $\kappa(\phi)$.
The  ``diffusion" coefficient $D$ for our one-dimensional  ballistic channel of length $L/2$ is  temperature independent and 
can be expressed in terms of the time of flight  $\tau_f=L/(2v_F)$ as $D=v_F^2 \tau_f =v_F L/2$, 
see Sect. III.2 in \cite{alhassid}.

The disk CFT partition function corresponding to a FQH droplet with
 filling factor $\nu_H=n_H/d_H$ can be written in 
general as a sum of products of   Luttinger-liquid partition functions, with interaction parameter $g=n_H d_H$,
representing the $u(1)$ charge,  
and neutral partition functions $\ch_{\Lambda'}(\tau)$ \cite{NPB-PF_k}
\beq \label{Z_CFT}
Z^{l}_{\Lambda}(\tau,\z)=\sum_{s=0}^{n_H-1} K_{l+s d_H}(\tau,n_H \z; n_H d_H) \, \ch_{\omega^s *\Lambda}(\tau),
\eeq
where $l$ (an integer defined $\mod d_H$) and $\Lambda$ denote respectively the $u(1)$ charge and neutral topological 
charge of the bulk quasiparticles, 
$\omega$ is the neutral topological charge of the electron operator \cite{NPB-PF_k}, $*$ represents  the fusion of two 
neutral topological charges and $\omega^s=\omega * \cdots * \omega$ is the s-fold fusion product \cite{CFT-book}. 
In most cases the neutral topological charge $\Lambda$ possesses a discrete $\Z_{n_H}$ quantum number $P(\Lambda)$ and 
satisfies the \textit{pairing rule}  $P(\Lambda) \equiv  l \mod n_H$.
The chiral Luttinger liquid grand canonical partition functions are explicitly \cite{NPB-PF_k}
\beq \label{K}
K_{l}(\tau,\z; m) = \frac{\mathrm{CZ}}{\eta(\tau)} \sum_{n=-\infty}^{\infty} q^{\frac{m}{2}\left(n+\frac{l}{m}\right)^2} 
\e^{2\pi i \z \left(n+\frac{l}{m}\right)},
\eeq
where $\eta(\tau)=q^{1/24}\prod_{n=1}^\infty (1-q^n)$ and the non-holomorphic factor $\mathrm{CZ}=\exp(-\pi\nu_H(\Im \z)^2/\Im\tau)$ is known as the Cappelli--Zemba factor \cite{cz}.
Because the statistics of the electron operator $ \theta/\pi=2\D=1/\nu_H +\theta(\omega)$ must be an odd integer
its neutral topological charge $\omega$ is always non-trivial when $n_H>1$ and so are the neutral characters 
$\ch_{\Lambda'}(\tau)$. 
For example, the neutral component of the electron for the $\Z_3$ Read--Rezayi (RR) FQH state  \cite{rr}
with $\nu_H=12/5$ is the parafermion field $\psi_1$ with $\D(\omega)=2/3$ \cite{NPB2001}.
Although the neutral partition $\ch_{\Lambda'}(\tau)$ are independent of $\z$ they  do change the flux periodicity 
of the conductance peaks because of their contributions to the 
electron energies.   For the numerical calculations below we use the following property  \cite{NPB2001}
\beq \label{shift}
K_l(\tau, n_H(\z+\phi\tau) ; n_H d_H) = K_{l+n_H\phi} (\tau,n_H\z; n_H d_H)
\eeq 
and set $\z=0$.
Since the electric charges of the excitations are given by the numbers multiplying $2\pi i \z$ in (\ref{Z_CFT})
and (\ref{K})
it is easy to see that the number of electrons, corresponding to excitations  with quantum numbers $n$
and $l$  (the same as in (\ref{K}) for $m=n_H d_H$),  in the right-hand-side of (\ref{shift}) is 
$N=n_H n+l/d_H + \nu_H \phi$ so that 
the addition of AB flux $\phi$, according to (\ref{phi}), changes the number of 
electrons by  $\nu_H \phi$ as it should be in order for the universal quantum Hall charge--flux relation 
(see Eq.~(3) in \cite{NPB-PF_k}) to hold. 
This gives the physical interpretation of the flux-threading transformation (\ref{phi}).
It also follows from (\ref{shift}),  (\ref{phi}), (\ref{Z_CFT})  and (\ref{K}) that the flux period is at most $\D\phi=d_H$
for any FQH state, because of the index-periodicity of the $K$ functions: 
$K_{l+n_H d_H}(\tau, n_H\z ; n_H d_H) = K_{l} (\tau,n_H\z; n_H d_H)$.
\begin{figure}[htb]
\centering
\includegraphics[bb=50 40 720 550,width=8.7cm,clip]{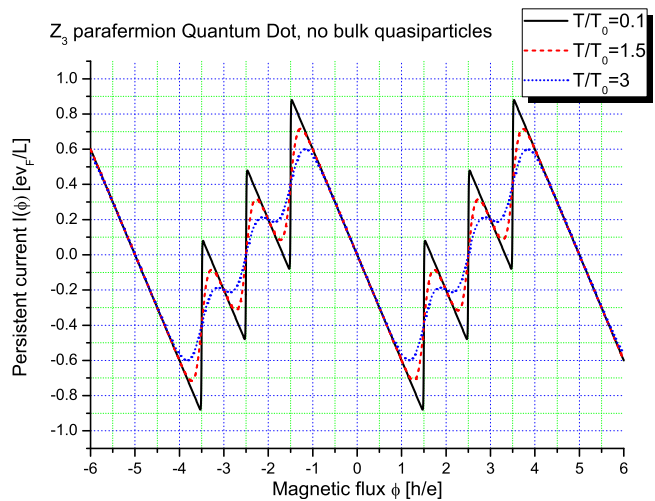}
\caption{(color online) Persistent current for a CB island which is in the $\Z_3$ parafermion FQH state without 
bulk quasiparticles.}
\label{fig:PF3-current}
\end{figure}

We consider the case without bulk-edge relaxation  
which means that the electron arriving at the edge
of the CB island moves fast enough from the left QPC to the right one without being able to fuse with bulk
quasiparticles. 
Under the assumption that the velocities of the charged and neutral modes are the same we plot in 
Fig.~\ref{fig:PF3-current} the persistent current $I(T,\phi)$,  as a function of the AB flux,  
for the $\Z_3$  RR state corresponding to $n_H=3$, $d_H=5$ and the neutral partition 
functions, $\ch_0\equiv \mathrm{Ch}_{00}$, $\ch_{\omega} \equiv \mathrm{Ch}_{01} $, 
$\ch_{\omega^2} \equiv \mathrm{Ch}_{02}$ corresponding 
to $l=0$, $\Lambda=0$ in (\ref{Z_CFT}) (i.e., no bulk quasiparticles), have been taken from Eq. (4.14) in \cite{NPB2001}.
If the velocities are  different then the flux periods can be calculated in a similar way.
We see  from Fig.~\ref{fig:PF3-current}  that the tunneling of a single electron into the edge 
of the CB island  corresponds to a jump, 
of universal size approaching $ev_F/L$ at zero temperature, in the persistent current and that $\D \phi = 5$.
\begin{figure}[htb]
\centering
\includegraphics[bb=60 10 750 600,width=8.7cm,clip]{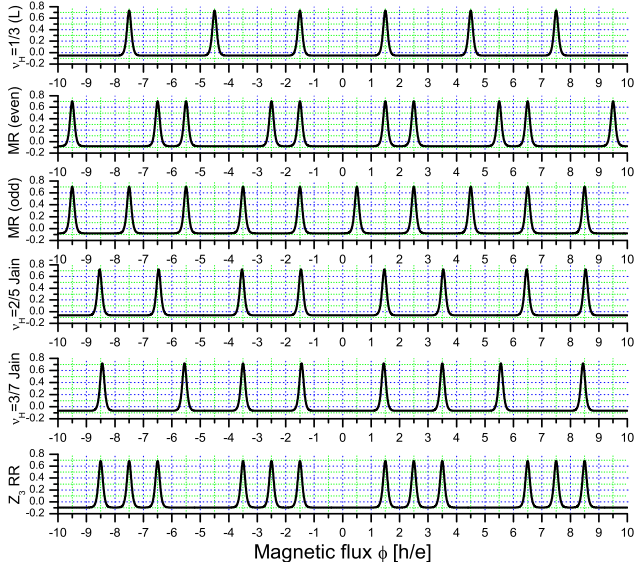}
\caption{Differential magnetic susceptibility $\kappa(\phi)$, at temperature $T/T_0=1$, in units 
$ \frac{e^2}{h}\frac{2\pi v_F}{L}$ for various FQH states with no bulk quasiparticles, except for the 
third plot---MR (odd)---which is of the MR state with a single bulk quasiparticle.}
\label{fig:CB-peaks}
\end{figure}
The corresponding peaks of the differential magnetic susceptibility are shown in the last plot of
Fig.~\ref{fig:CB-peaks}, from which we see that the peaks are clustered in triples separated by $\D\phi_1=1$ 
inside of the cluster and by $\D\phi_2=3$ between the clusters. 
The other plots in Fig.~\ref{fig:CB-peaks} are constructed from the following partition functions (\ref{Z_CFT}):
$\nu_H=1/3$ (L) is the  Laughlin state ($n_H=1$, $d_H=3$, no neutral partition functions); MR (even) is
the Moore--Read (MR) FQH state \cite{mr},  $n_H=2$, $d_H=4$ with neutral partition functions, for $l=0$, $\Lambda=0$ 
in (\ref{Z_CFT}), $\ch_0$ and $\ch_{\omega}=\ch_{1/2}$ taken from Eq.~(7) in \cite{5-2};  MR (odd) is the MR state 
with one quasiparticle in the bulk and for  $l=1$, $\Lambda=\s$, where $\s$ is the Ising anyon,  with neutral partition 
functions $\ch_{\Lambda}=\ch_{\omega*\Lambda}=\ch_{1/16}$ taken from Eq.~(7) in \cite{5-2}---this CB peaks pattern is identical 
with that for the  $g=1/2$ Luttinger liquid--this change of the total flux period from $4$ to $2$ is the even--odd effect
in the MR state\cite{stern-halperin-5-2}; 
the fourth plot corresponds to the hierarchical $\nu_H=2/5$ Jain FQH state with $n_H=2$, $d_H=5$ and
neutral  partition functions, for $l=\Lambda=0$,  $\ch_0=K_0(\tau,0;2)$ and  $\ch_{\omega}=K_1(\tau,0;2)$;
the fifth plot is of the $\nu_H=3/7$ Jain state, which corresponds 
to $n_H=3$, $d_H=7$ in Eq.~(\ref{Z_CFT}) with $l=\Lambda=0$ and  neutral partition functions $\ch_0=\ch_{(0,0)}$, 
$\ch_{\omega}=\ch_{(1,0)}$ and $\ch_{\omega^2}=\ch_{(0,1)}$ taken as the characters \cite{CFT-book} of the vacuum and 
fundamental representations of the current algebra $\widehat{su(3)}_1$. 
The flux spacing of the peaks in Fig.~\ref{fig:CB-peaks} are in perfect agreement with those of  the CB peaks
obtained earlier in the literature for zero temperature
\cite{stern-halperin-5-2,stern-CB-RR,stern-CB-RR-PRB,nayak-doppel-CB,CB,cappelli-viola-zemba}
by analyzing the flux dependence of the electron energies at the CB resonances.
However, our results also allow us to estimate the shape, height and width of the CB peaks 
at finite temperature, which is important for the experiments.
\begin{figure}[htb]
\centering
\includegraphics[bb=25 22 541 430,width=7cm,clip]{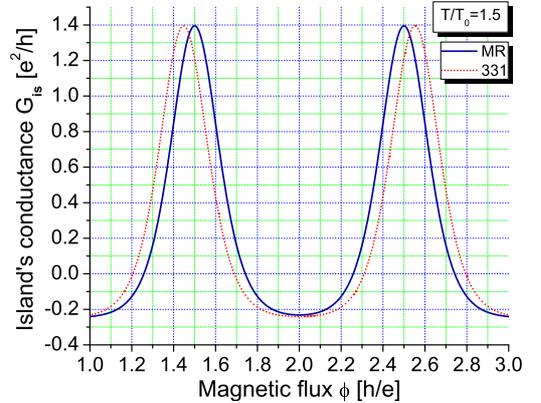}
\caption{(color online) Island's conductance for the Moore--Read and 331 FQH states
(without bulk quasiparticles).}
\label{fig:K-MR-331}
\end{figure}
 One interesting feature of the CB peaks, which demonstrates the advantage of the CFT approach, 
is that they may become displaced at finite $T$, in states such as the 331, $\nu_H=2/5$ and 
$\nu_H=3/7$ Jain states,  due to electron multiplicities in the neutral sector. 
 Consider, e.g., the 331 (Halperin) FQH state, which  corresponds to  $n_H=2$, $d_H=4$ and
 neutral partition functions $\ch_0=K_{0}(\tau,0;4)$  and $\ch_{\omega}=K_2(\tau,0;4)$. It has been demonstrated 
earlier that the CB peaks patterns of the 331 and MR state are indistinguishable at $T=0$ 
and the situation is similar for other FQH states with different topological orders \cite{nayak-doppel-CB}. 
When the temperature increases, the CB peaks in the 331 state are displaced in such a way that the
short period $\D \phi_1=1 $ increases (by approximately $0.1$  h/e in Fig.~\ref{fig:K-MR-331}), while the long one
 $\D \phi_2=3 $ decreases keeping the total periodicity $\D \phi = 4$ unchanged. 
This can be explained as follows---due to neutral multiplicities, $m=2$ in this case, the neutral character of the 
electron sector (without bulk quasiparticles), for $T\ll T_0$, is $\ch_{\omega} (\tau) \simeq m q^{\D_{0} }= q^{{\D'}_{0} }$
where $\D_{0}$ is the neutral CFT dimension of the electron and  
\beq\label{neutral}
{\D'}_{0} =\D_{0} - \frac{1}{2\pi^2}\frac{T}{T_0}\ln m.
\eeq 
Thus, increasing $T$ effectively lowers the neutral energy of the electron and therefore displaces the CB peaks,
which appear at flux positions at which two parabolas, shifted in the vertical direction by the neutral electron 
energy, cross (see e.g., Fig.~3 in \cite{cappelli-viola-zemba}).
We see from our Fig.~\ref{fig:K-MR-331}, that the  finite-temperature CB peak patterns of the MR and 331 states 
are not identical and this temperature dependence of the CB peak's periods could in principle be used to 
distinguish between them.
 
\begin{figure}[htb]
\centering
\includegraphics[bb=20 15 540 430, width=7cm,clip]{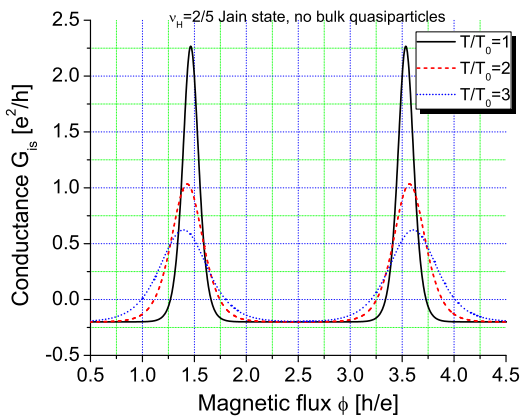}
\caption{(color online) Peak asymmetry for the $\nu=2/5$ Jain FQH state due to the neutral-sector  multiplicities. 
This picture is reproduced periodically along the $\phi$ axis with period  $\D\phi=5$.}
\label{fig:Jain25}
\end{figure}

Similar asymmetry and displacement of the CB peaks could be seen in the $\nu=2/5$ Jain state,
where the neutral electron multiplicity is $m=2$ again \cite{cappelli-viola-zemba}, see Fig.~\ref{fig:Jain25}. 
 The distances between the peaks are 
$\D \phi_1 =2$,   $\D \phi_2 = 3$ giving total periodicity $\D \phi =5$.
When temperature increases the short period $\D \phi_1$ tends to increase
while   $\D \phi_2 $ decreases keeping the total periodicity the same. 
Also, in the $\nu_H=3/7$ Jain FQH state 
the tendency is again for the short period $\D\phi_1=2$ to increase, while the long period 
$\D\phi_2=3$ to decrease when increasing the temperature, and this  is  again attributed 
to the neutral multiplicities \cite{cappelli-viola-zemba}.

\begin{figure}[htb]
\centering
\includegraphics[bb=35 15 540 430,width=6.5cm,clip]{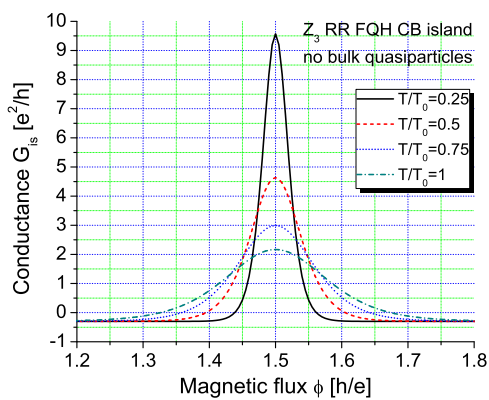}
\caption{(color online) Broadening of the CB peaks in the $\Z_3$ parafermion (RR) FQH droplet without 
bulk quasiparticles.}
\label{fig:broadening}
\end{figure}

For the $\Z_3$ RR state the positions of the peaks and the distances between them do not change with 
temperature, but the peaks decrease and broaden, see Fig.~\ref{fig:broadening}.
Eq.~(\ref{Z_CFT}) allows us to derive an analytical estimate of
the height, width and shape of the CB peaks at low temperature. 
Keeping only the leading terms in (\ref{Z_CFT}) for $q=\exp(-\D\epsilon/k_BT) \to 0$, 
which for $l=\Lambda=0$ are the terms with $s=0,\pm 1 (\mod n_H)$ in (\ref{Z_CFT}) and $n=0$ in (\ref{K}), 
it can be shown that  the partition function (\ref{Z_CFT}) has the following low-temperature approximation
(no bulk quasiparticles) for $|\phi-\D| <1/2$
\beq\label{Z-low}
Z(T,\phi) \mathop{\simeq}_{T\ll T_0} q^{\nu_H \frac{\phi^2}{2}} 
\left[ 1+2 q^{\D}\cosh\left( \frac{\D\epsilon}{k_B T} \phi\right)\right],
\eeq
where $\D=1/(2\nu_H) + {\D}_0$ is the total CFT dimension of the electron. 
Then the CB conductance (\ref{G_CB}) for the case without neutral multiplicities 
has a universal peak shape given by
\beq \label{shape}
G_{\mathrm{CB}} \mathop{\simeq}_{T\ll T_0}  \frac{g_L g_R}{g_L+g_R} \frac{e^2}{h}\left(-\frac{1}{2} \right)
\frac{\partial}{ \partial\phi} \left( \frac{1}{1+\e^{\frac{\D\epsilon}{k_B T} (\phi-\D)} }\right) 
\eeq
and therefore the peak's height can be estimated to be
\beq\label{peak}
G_{\mathrm{CB}}^{\mathrm{peak}} \mathop{\simeq}_{T\ll T_0} 
\frac{g_L g_R}{g_L+g_R} \frac{e^2}{h} \frac{\D\epsilon}{8k_B T},
\eeq
which is consistent with previous estimates (see Eqs.~(24) in \cite{alhassid}, Eq.~(134) in \cite{PRept-CB} and
Eq.~(1) in \cite{furusaki}).
Notice that (\ref{peak}) has crucial implications for the experimental observability of the CB peaks at low 
temperatures. 
{The dimensionless tunneling conductances $g_{L,R}$ for the left and right QPC's, defined in (\ref{G_CB}), 
which appear  in Eq.~(\ref{peak})  depend on temperature \cite{furusaki}
as $g_{L,R}\propto T^{2\D_{L,R}+2\D_{QD}-2}= T^{4\D -2}$, where $\D_{L,R}$ is the CFT dimension of the electron in the 
Left,Right lead, while $\D_{QD}$ is the CFT dimension on the quantum dot  and we have chosen  
$\D_{L}=\D_{R}=\D_{QD}=\D$. According to (\ref{peak}) the island conductance's peak is 
$G_{\mathrm{is}} \propto T^{-1}$ and therefore the total CB conductance vanishes at zero temperature
\beq
G_{\mathrm{CB}}^{\mathrm{peak}} \propto T^{4\D-3}, \quad T\ll T_0
\eeq
 for most  FQH liquids since the electron dimension (which must be half-integer) is $\D \geq 3/2$. 
The coefficient of proportionality depends on the bare tunneling 
amplitudes of the left and right QPC's which do not renormalize because the electron tunneling is irrelevant 
in the sense of the RG flow and therefore the strong-reflection regime of the Fabry--Perot interferometer, shown 
in Fig.~\ref{fig:FPI-CB}, is a 
stable RG fixed point. Notice that the conductance scaling with temperature, which we use here corresponds to
Sect.~IV in Ref.~\cite{furusaki} and has to be distinguished from the results of Sect.~III there, where the quantum dot
contribution has not been taken into account correctly.}

Substituting (\ref{neutral}) in (\ref{Z-low}) for the 331 and $\nu_H=2/5$ Jain states we obtain that 
the short flux period increases as a function of the temperature as 
\beq
\D\phi_1 (T)=\D\phi_1(0) +  \frac{1}{\pi^2}\frac{T}{T_0}\ln m,
\eeq
where the neutral electron multiplicity is $m=2$.

{As can be seen from Eq.~\ref{shape} and Figs.~\ref{fig:K-MR-331}, \ref{fig:Jain25} and \ref{fig:broadening},
the (sequential-tunneling) conductance decreases exponentially away from the CB peak's centers and becomes negative in the 
CB valleys, however, the co-tunneling conductance decays only algebraically \cite{furusaki}, hence is the main transport mechanism
in the CB valleys, and the total conductance remains positive}.

If we define the width $\delta\phi = |\phi_2-\phi_1|$ of a CB peak  as the 
difference between the two   flux positions $\phi_1$ and $\phi_2$, around the peak, at which 
$G_{\mathrm{is}}(\phi_{1,2})=0$, resp.,  $\kappa(\phi_{1,2})=0$, then using 
(\ref{Z-low}) we can prove that the peaks' width is not simply proportional to the temperature---instead, there is a 
logarithmic correction  
\beq \label{width}
\delta \phi (T)\mathop{\simeq}_{T < 3T_0} \frac{1}{\pi^2} \left(\frac{T}{T_0}\right) 
\left[ \ln\left(\frac{2\pi^2}{\nu_H}\right) -\ln\left(\frac{T}{T_0}\right)\right].
\eeq
{However, as we discussed above, when the conductance begins to decrease, due to deviation of the flux from the CB peak's 
center, there is a crossover to co-tunneling regime.
Therefore we could define the CB peak's width at level $\sigma$, such that the conductance at the points $\phi_{1,2}$ defining the 
width is  $G_{\mathrm{CB}}(\phi_{1,2})=\s G_{\mathrm{CB}}^{\mathrm{peak}}$,  where sequential tunneling is still the dominating process.
Then, ignoring a logarithmic term similar to that in (\ref{width}), we find that the width of the CB peak at level $\s$ increases 
linearly with temperature
\beqa
\left. \delta \phi(T)\right|_{\sigma} &=& \frac{\alpha}{2\pi^2}  \left(\frac{T}{T_0}\right) , \quad \mathrm{where} \quad 
\frac{1}{2} \leq\sigma <1 \ \mathrm{and}\nn
\alpha&=&-2\ln \left[ \left( \frac{2}{\sigma}-1\right)  -\sqrt{ \left( \frac{2}{\sigma}-1\right)^2-1} \right] . \nonumber
\eeqa
Notice that this estimate of the CB peak's width is independent of the filling factor $\nu_H$ because it is derived 
from Eq.~(\ref{shape}) which is universal and independent of $\nu_H$.
}

In conclusion, we showed that the CB conductance peaks, {for a Fabry--Perot interferometer realized in a FQH bar}, 
could be computed at finite temperature in terms of the differential 
magnetic susceptibility derived within the {equilibrium} effective CFT for the FQH edge states.
{Using the known chiral partition functions for a number of FQH states we derived the positions of the CB peaks, 
their spacing, universal shape, height and width. While the CB peak spacing coincides with previous results in the literature,
the new estimates for the peak's height and width at finite temperature could give important hints about the experimental 
observation of the CB. In addition, the asymmetry and displacement of the CB peaks positions at finite temperature could 
give important information about the neutral multiplicities in the corresponding FQH states, eventually providing signatures 
to distinguish different FQH states with similar zero-temperature CB patterns. }

\textit{Acknowledgments:} I am grateful to Reinhold Egger for useful discussions.
 The author has been supported by the Alexander von Humboldt Foundation.
This work has been partially supported by  the BG-NSF under Contract No. DO 02-257.
\bibliography{Z_k,my,FQHE}

\end{document}